\documentclass[prl,twocolumn,amsmath,amssymb]{revtex4}
\usepackage{amssymb}

\usepackage{amsmath}
\usepackage{epsf}
\usepackage{graphicx}

\begin{document}


\title{Resonance mode in $B_{1g}$ Raman scattering -- a way to
  distinguish between spin-fluctuation and phonon-mediated
 $d-$wave superconductivity}

\author{A. V. Chubukov$^1$, T. P. Devereaux$^2$ and M. V. Klein$^3$}

\affiliation{$^1$
Department of Physics, University of Wisconsin, Madison, WI 53706}
\affiliation{$^2$Department of Physics, University of Waterloo,
Waterloo, Ontario Canada N2L 3G1}
\affiliation{$^3$Department of Physics, University of Illinois,
Urbana, IL 61801}
\date{\today}

\begin{abstract}
We argue that Raman scattering in $B_{1g}$ symmetry allows one to distinguish between
phonon-mediated and magnetically-mediated $d-$wave superconductivity.
In spin  mediated superconductors, $B_{1g}$ Raman intensity develops a
 resonance at a frequency  $\omega_{res} < 2\Delta_{max}$,
 whose origin is similar to a neutron resonance. In phonon-mediated
 $d-$wave superconductors, such a resonance does not develop. Several
extensions of the argument are presented.
\end{abstract}


\maketitle

This paper is devoted to the analysis of whether there exists an observable that would distinguish between magnetically mediated and phonon-mediated
$d-$wave superconductivity. This subject is motivated by the cuprates
for which recent measurements,
particularly the observation of the
kink in the quasiparticle dispersion~\cite{kink},
has revived the discussion of whether
the pairing in the cuprates may be due to phonons rather than
spin fluctuations.  Both electron-phonon~\cite{tom_ph}  and spin-fermion interactions~\cite{mike}
were advanced to explain the features in the quasiparticle dispersion. Thus,
to truly distinguish between the two scenarios, one needs to identify an observable for which phonon and spin-fluctuation pairings yield qualitatively
different results.

In this communication, we argue that
Raman scattering in $B_{1g}$ symmetry  is such a probe.
We show that in spin  mediated $d-$wave superconductors, the $B_{1g}$ Raman intensity develops a
resonance at a frequency  $\omega_{res} < 2\Delta_{max}$, while
in  phonon-mediated $d-$wave superconductors, such
a resonance does not develop.
The resonance is excitonic in origin, and is similar to the excitonic
 resonance
in the spin susceptibility of a $d-$wave superconductor~\cite{neutrons}.
 The only difference is that the resonance in $B_{1g}$ Raman scattering comes from fermions all around the Fermi surface and
has a finite intrinsic width because of the nodes,
whereas the  resonance term in the spin susceptibility
comes only from fermions in the antinodal regions
 and is a true, $\delta-$functional bound state.

To make the argumentation straightforward, we first
consider the  simplest case of $S=1/2$ fermions
 interacting via a  static potential $V^{pair} (k)$
\begin{equation}
\label{Hint}
  {\cal H}_{\rm int} = - \sum_{\bf{q},\bf{k},\bf{p}}
  \psi_{\bf{k},\alpha}^{\dagger} \psi_{\bf{p} + {\bf q}, \beta}^{\dagger}
  ~V^{pair}_{\alpha \beta, \gamma \delta} (\bf{k} - {\bf p})~
\psi_{\bf{p} \gamma} \psi_{\bf{k} + {\bf q}, \delta}.
\end{equation}

In Eq.(1) summation over spin indices $\alpha, \beta, \gamma$ and $\delta$ is understood.
As $B_{1g}$ vertex has the same $d-$wave form as the pairing gap,
we further approximate $V^{pair}$ by its $d-$wave component
$V^{pair} ({\bf k} - {\bf p}) \propto d_k d_p$,
where $d_k = [\cos (k_{x}a)-\cos (k_{y}a)]/2$.

The effective interaction $V^{pair}_{\alpha \beta, \gamma \delta}
 ({\bf k} - {\bf p})$ may be due to spin fluctuations
 or to phonons.
For  spin mediated interaction,
 $V^{pair}_{\alpha\beta,\gamma\delta} ({\bf k} - {\bf p})
= V_{spin} d_k d_p {\boldsymbol \sigma}_{\alpha \gamma}\cdot {\boldsymbol \sigma}_{\beta \delta}$ where $\boldsymbol \sigma$ are Pauli matrices.
For phonon-mediated interaction
$V^{pair}_{\alpha\beta,\gamma\delta} ({\bf k} - {\bf p})
= V_{ph} d_k d_p \delta_{\alpha \gamma} \delta_{\beta \delta}$.
We assume that both interactions lead to a
$d-$wave pairing, and study the consequences
 for the Raman intensity.
>From this perspective, our results
are equally applicable if phonons are replaced by charge density waves~\cite{grili}.

Our first observation is that the signs of $V_{spin}$ and $V_{charge}$
must be different, if they both lead to an attraction in a $d-$wave channel.
Indeed, substituting effective interctions into the diagrammatic expression
 for the $d-$wave, spin-singlet pairing vertex
$\psi_{\bf{k},\alpha}^{\dagger}\Phi_d (k)_{\alpha\beta} \psi_{\bf{-k}, \beta}^{\dagger}$,
where $\Phi_d (k)_{\alpha\beta} = \Phi d_k \sigma^y_{\alpha\beta}$,
and using
$\sigma^y_{\alpha\beta} \delta_{\alpha\gamma} \delta_{\beta\delta} = \sigma^y_{\gamma\delta},~~\sigma^y_{\alpha\beta}
{\boldsymbol \sigma}_{\alpha\gamma}\cdot
{\boldsymbol  \sigma}_{\beta\delta} = -3 \sigma^y_{\gamma\delta}$,
we obtain that $\Phi$ is related to the
 bare vertex $\Phi_0$ as
\begin{equation}
\Phi  = \frac{\Phi_0}{1 + 3 A V_{spin} }; ~~~~\Phi = \frac{\Phi_0}{1 - A V_{ph}}\label{4}
\end{equation}
for spin-mediate or phonon-mediated interaction, respectfully.
A positive $A \propto |\log{\omega_{c}}|$ is a conventional
 logarithmic cut-off factor.
To get an attraction, one then obviously needs  $V_{ph}$ to be  positive,
$V_{spin}$ to be negative. This is the case when phonon mediated interaction is peaked at small momenta, and spin-mediated nteraction is peaked at momenta near $(\pi,pi)$~\cite{spin_fl,phon}.

Suppose next that the system is  a $d-$wave superconductor, either due to phonons or due to spin fluctuations. Consider how this affects the
Raman response in the $B_{1g}$ channel.
The $B_{1g}$ Raman vertex $\Gamma_{\alpha\beta} (k)$  has $d-$wave momentum dependence and is a spin scalar:
 $\Gamma_{\alpha \beta} (k) = \Gamma d_k \delta_{\alpha\beta}$.
For a BCS superconductor without vertex or self energy corrections, the $B_{1g}$ Raman intensity
$I_{B_{1g}} (\Omega) =  \Gamma^2 Im \chi_0 (\Omega)$
and $\chi_{0} (\omega)$  is  the imaginary part of the
 particle-hole bubble with two $d$-wave vertices~(see e.g. Ref.\cite{miles_1}).  In the normal state, $\chi_0 (\Omega)$ vanishes, as it should as
there is no low-energy phase-space available for scattering with ${\bf q}=0$.
In the superconducting state, light can break Cooper pairs with ${\bf q}=0$ and
$\chi_0 (\Omega)$ is given by  Tsuneto
function weighted with $d^2_k$ \cite{miles_1,tom_w3}
The imaginary part of $\chi_0 (\Omega)$
 scales as $\Omega^3$ at small frequencies~\cite{tom_w3}, and
 diverges logarithmically as $\Omega$ approaches $\pm 2\Delta$: Im$\chi_0 (\Omega) \propto
\log [\Delta/\sqrt{\Omega^2 - 4 \Delta^2}]$.

\begin{figure}
\includegraphics[width=0.8\columnwidth]{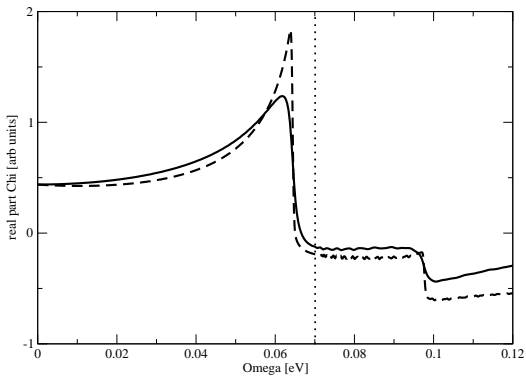}
\caption{The real part of the $B_{1g}$ Raman response in the superconducting state Re$\chi_{0}$ for the case of
small fermionic damping (solid line) and no fermionic damping (dashed line). The vertical dotted line
marks $2\Delta_{0}$. A small change between $2\Delta_0$ and the frequency where $Re \chi_0 (\Omega)$ is peaked
is due to the fact that the maximum value of the gap along the Fermi surface is slighly smaller than $\Delta_0$.}
\label{real_chi}
\end{figure}

 The corresponding real part
at small frequencies varies as Re$\chi_0 (\Omega) = N_{0}[1 + O(\Omega^2/\Delta^2)]$, with
$N_{0}$ the density of states at the Fermi level, and keeps increasing up to $2\Delta$ before discontinuously
jumping across zero at a frequency of twice the maximal gap on the Fermi surface.
Another milder
jump occurs at twice the energy at the van Hove points at $(\pi,0)$ and related points in the Brillouin
zone. We plot Re$\chi_0 (\Omega)$ in Fig \ref{real_chi}, using
$\Delta_{0}=35$ meV and the band structure
$\epsilon_{k}$ given by Eschrig and Norman\cite{Norman} to fit the angle-resolved photoemission (ARPES)
on optimally doped Bi$_{2}$Sr$_{2}$CaCu$_{2}$O$_{8+x}$
(Bi-2212).
Although the generic behavior changes near $2\Delta_{0}$ if damping
(phenomenologically represented by $\delta$) is increased,
the rapid rise and fall of the real part near $2\Delta$ is preserved.

Consider now what happens to the Raman response when we add $V_{pair}$. This interaction leads to two effects -
self-energy renormalization of the fermions in the particle-hole bubble, and the
renormalization of the Raman vertex. The self-energy renormalization does not distinguish
qualitatively between phonons and spin fluctuations, and still preserves the peak
$I_{B_{1g}} (\Omega) $ at $\Omega=2\Delta$. The
 renormalization of the $B_{1g}$ vertex  is more relevant.
To understand its role we first observe that there is no spin-induced sign change between vertex renormalization due to phonons and due to spin fluctuations.
Indeed, convoluting the spin dependence of
$V^{pair}_{\alpha\beta,\gamma\delta}$ with
$\delta_{\beta\gamma}$  of the Raman vertex $\Gamma_(\beta\gamma) (k)$,
we find that for magnetic interaction, the summation over spin indices yields
$
\delta_{\beta\gamma} {\bf \sigma}_{\alpha \gamma} {\bf \sigma}_{\beta \delta}
 = \sum_i (\sigma^2_{\alpha \delta})_i = 3 \delta_{\alpha \delta}$,
while for phonons
$
\delta_{\beta\gamma} \delta_{\alpha \gamma} \delta_{\beta \delta} = \delta_{\alpha\delta}$.
As a result, in both cases, the spin configuration of the Raman vertex
is reproduced, and the sign is the same in both cases.
Summing up the vertex correction diagrams in the ladder approximation, we
obtain
\begin{eqnarray}
&&I^{ph}_{B_{1g}} (\Omega) =
\Gamma^2 \frac{\chi^{\prime \prime}_0 (\Omega)}{(1 +
\frac{1}{2}~V_{ph} \chi^\prime_0 (\Omega))^2 + \left(\frac{1}{2}V_{ph} \chi^{\prime \prime}_0 (\Omega)\right)^2} \nonumber \\
&&I^{spin}_{B_{1g}} (\Omega) =
\Gamma^2 \frac{\chi^{\prime \prime}_0 (\Omega)}{(1 +
\frac{3}{2} V_{spin} \chi^\prime_0 (\Omega))^2 + \left(\frac{3}{2} V_{spin} \chi^{\prime \prime}_0 (\Omega)\right)^2}
\label{9}
\end{eqnarray}
We defined $\chi^\prime = Re \chi, ~\chi^{\prime \prime} = Im \chi$.
 We now recall that
$V_{spin}$ must be negative, and $V_{ph}$ must be positive, if each of them
gives rise to a $d-$wave pairing.
Then, the sign of the vertex renormalization in (\ref{9}) {\it is different
 for phonons and spin fluctuations}.
As mentioned above, Im $\chi_0 (\Omega)$
in the superconducting state  is quite small except for near $2\Delta$.
The renormalization of the Raman vertex at $\Omega <2\Delta$ then predominantly comes from Re$\chi_0$.

\begin{figure}
\includegraphics[width=0.8\columnwidth]{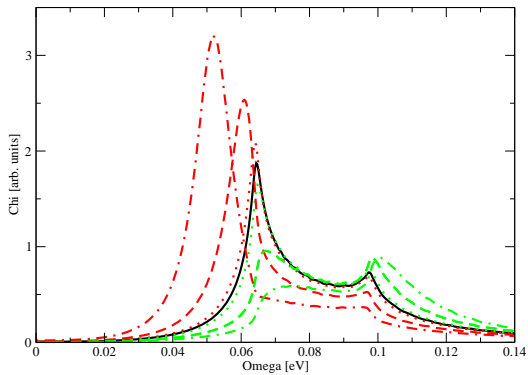}
\caption{The full Raman response plotted for $VN_{0}=0$ (black solid line), and
$\pm 0.01,0.1,0.5$ (dotted, dashed, dashed-dotted lines, respectively) for phonon-mediated
(green lines) and spin-mediated (red), respectively.}
\label{full}
\end{figure}

Since Re$\chi_0 (\Omega)$ is positive, vertex renormalization due to phonons
reduces the Raman vertex at small frequencies and only slightly shifts up
the peak which remains close to $2\Delta$. On the other hand, if the $d-$wave
interaction is magnetic in origin, $V_{spin} \chi_0^\prime (\Omega) <0$,
and for strong enough $V_{spin}$, there exists a frequency $\Omega_{res} <2\Delta$
at which $(3/2) V_{spin} \chi_0^\prime (\Omega) =-1$.
At this frequency, $I^{spin}_{B_{1g}}$ has a peak.
Because $\chi^{\prime \prime}_0 (\Omega)$ is nonzero at any $\Omega >0$, the peak is not infinitely sharp as in an
$s-$wave superconductor\cite{Fred}.
However, since $\chi^{\prime \prime}_0 \propto \Omega^3$  at small frequencies, the
width of the peak is small if  $\Omega_{res}$ is substantially smaller than $2\Delta$.
We see therefore that
for spin-mediated $d-$wave pairing, the $B_{1g}$ Raman intensity develops a resonance.
The resonance for spin-mediated pairing was
discovered in Ref~\cite{cmb}, although its origin was not discussed in detail there. We plot in Figure \ref{full} the full $B_{1g}$ Raman response for $T=0$ in the superconducting state for
different values of the interaction $N_{0}V$ using the parameters shown for Figure (\ref{real_chi}). In the absence of
interactions $(V=0)$ the Raman response rises as $\Omega^{3}$ and
has a clear peak at twice the gap and another smaller peak at twice the van Hove energy. For magnetic interactions
$V<0$ the low energy peak sharpens and moves to lower frequency as the resonance develops and steals spectral weight
from the $2\Delta_{0}$ feature. In agreement with Ref. (\cite{cmb}), two separate peaks
do not develop, but because of resonance, the
original  $2\Delta$ peak progressively shifts down from $2\Delta$ as the interaction increases. Conversely,
for phononic interactions $V>0$,
the peak renormalizes upwards from the final-state interactions and weakens for
stronger interactions. No low energy resonance develops.

The resonance in the $B_{1g}$ Raman response is in many respects
similar to the resonance in the spin susceptibility in a $d-$wave
superconductor~\cite{neutrons}. In both cases, the $d-$wave
symmetry of the gap is crucial, and the resonance emerges due to
residual attraction between fermions and spin-fluctuations in a
$d-$wave superconductor. The only difference between the two
resonances is that the neutron resonance is virtually a bound
state in the sense that it is infinitesimally narrow, and does
not require  coupling to be above a threshold because the real
part of the bare spin susceptibility $\chi_s (Q, \Omega)$, where
$Q = (\pi,\pi)$, evolves between $0$ and infinity at $0 < \Omega
< 2\Delta$, where Im $\chi_s (Q, \Omega) =0$. This comes about
because the spin susceptibility at momentum $Q$ is determined by
fermions near hot spots, where $k$ and $k+Q$ are both near the
Fermi surface. The hot spots are generally located away from the
nodes, hence in the superconducting state hot fermions are fully
gaped, and a spin fluctuation needs a finite energy to be able to
decay into a particle-hole pair. The Raman resonance, on the
other hand, is a $Q=0$ probe. It therefore involves fermions from
the entire Fermi surface, including nodal regions. The nodal
fermions account for a nonzero $Im \chi_0 (\Omega)$ at any finite
$\Omega$ and therefore produce an  intrinsic width of the Raman
resonance peak. The interaction then should be above the
threshold for the resonance to become visible.

We now extend the above analysis in several different directions.
First, when $V_{spin}$ and $V_{ph}$ are both nonzero, the
resonance condition  becomes $(1/2) V_{eff} \chi_0^\prime
(\Omega) =- 1$, with $V_{eff}=3 V_{spin} + V_{ph}$. We remind the
reader that $V_{spin}<0$ and $V_{ph}>0$ both favor $d-$pairing,
and thus they compete to determine $V_{eff}$. Thus referring back
to Figure (\ref{full}) the full $B_{1g}$ Raman response is a
function of the net pairing interactions $V_{eff}$, and the
presence of the resonance requires $3V_{spin}>\mid
V_{phonon}\mid$ if $d-$pairing occurs via both channels.

Second, the analysis can be extended to the case where the pairing
interaction and the Raman vertex have the same symmetry, but are not identical.
If $V^{pair} ({\bf k} - {\bf p})
\propto \phi_{\bf k} \phi_{\bf p}$ and
the bare Raman vertex is $\Gamma_{\bf k} = \Gamma \gamma_{\bf k}$, where the functions $\phi_{\bf k}$ and $\gamma_{\bf k}$ belong to the same irreducible
representation different from fully symmetric $A_{1g}$, the RPA Raman
 intensity near the resonance at a temperature $T$ is given by
\begin{equation}
I_{\gamma_{\bf k}} (\Omega) \sim Im \left[\frac{\Delta^2 K_{10} K_{01}}{1 + 2\Delta^2 (3 V_{spin} + V_{ph})
K_{00}}\right]
\label{n_1}
\end{equation}
for spin and phonon mediated pairing, where
\begin{equation}
K_{n,m} (\Omega,T) = \int \frac{d^2 k}{(2\pi)^2}
 \frac{(\phi_{\bf k})^{4-n-m} \gamma^n_{\bf k} (\gamma^*_{\bf k})^m}{E_{\bf k}
(4E^2_{\bf k}- (\Omega+ i \delta)^2)}~~\tanh {\frac{E_{\bf k}}{2T}}.
\label{n_2}
\end{equation}
and $E^2_{\bf k} = \epsilon^2_k + \Delta^2 \phi^2_{\bf k}$.
One can verify that for $\phi_{k} = d_k$,
$K_{00} (\Omega, 0) = K_{10} (\Omega,0)= K_{01} (\Omega,0) =
(1/4\Delta_{0}^2) \chi_0 (\Omega)$, and (\ref{n_1}) coincides with (\ref{9}). Eq. (\ref{n_1}) however shows
that once $3 V_{spin} + V_{ph}$ is negative, the resonance occurs for any
pairing symmetry different from $A_{1g}$, for the Raman vertex for which
$K_{01}$ and $K_{10}$ are finite.

Third, we argue that for spin mediated pairing $d-$wave pairing, the resonance in the $B_{1g}$
channel survives even when the effective interaction
includes a dynamic term, e.g., the Landau damping, and vanishes at
high frequencies. The vanishing of $V^{pair} (k, \omega)$ at large $\omega$
may have a profound effect on the renormalization of the vertex
at zero momentum transfer as the real part of the vertex renormalization,
which accounts for the resonance in our case,
partly comes from fermions with high frequencies. Once the interaction
 vanishes at high frequencies, this part disappears, and it becomes an issue whether the remaining part  still has the same sign.

To address this issue, we  computed $V_{spin} \chi_0 (\Omega)$
in  Eliashberg theory which includes Landau damping. We obtained
the expression for the Raman vertex renormalization
$\Gamma_{B_{1g}} (\Omega) = \Gamma d_k/ (1 +J(\Omega))$ by first
evaluating the vertex correction in Matsubara frequencies, and
then analytically continuing to the real axis by introducing
double spectral representation.  We found  that $J(0) =0$, but
$J(\Omega) <0$ at $\Omega < 2\Delta$ still logarithmically diverges at
$2\Delta$,  i.e.,  vertex corrections, even without high frequency term, still
increase the Raman vertex and lead to a resonance  in $R(\Omega)$ at some $\Omega < 2\Delta$.

\begin{figure}
\includegraphics[width=0.8\columnwidth]{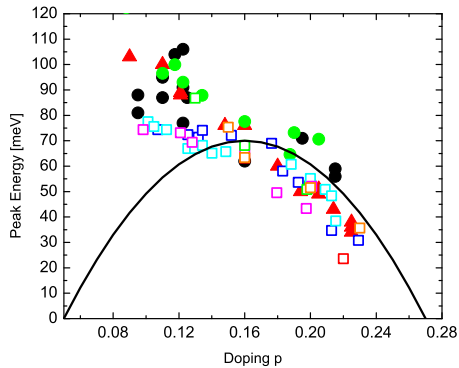}
\caption{The Raman data (open squares) from five different groups
(Ref. \protect\cite{overdoped} (red), Ref. \protect\cite{Blumberg}
(magenta), Ref. \protect\cite{Hackl} (orange), Ref.
\protect\cite{Sugai} (green), Ref. \protect\cite{Hewitt} (blue),
Ref. \protect\cite{Kendziora} (cyan), respectively), together with
the tunneling data (red triangles) from ~\protect\cite{Tunnel}
and ARPES data from ~\protect\cite{ARPES} (black circles are
derived from leading edge spectra while green circles are from
the peak position of the $\pi,0$ spectra). The doping $p$ was
determined
by the formula $p = 0.16 \pm
\sqrt{\frac{|T_{c}-T_{c}^{max}|}{82.6T_{c}^{max}}}$, and the
solid line plots 70 meV $\rm{\times T_{c}/T_{c}^{max}}$.}
\label{comparison}
\end{figure}

The only extensive set of data comparing $B_{1g}$ symmetry Raman
gap values with those from the single electron spectroscopies of
angle-resolved photoemission spectroscopy (ARPES) and tunneling
is found for the Bi$_{2}$Sr$_{2}$CaCu$_{2}$O$_{8+x}$ ("Bi-2212")
family. Values found by five groups for the Raman gap in terms of
the hole doping $p$ are shown in Fig \ref{comparison}. The Raman
gap values are compared in this figure
with those from tunneling and ARPES.
 Tunneling results are from Ref~\cite {Tunnel}
and represent peak-to-peak separations in positive and negative
biases.  ARPES results are from Ref~\cite {ARPES}. For completeness, we
 presented twice the gap $\Delta$ determined from two sets of
ARPES data - the position of the peak at $(0,\pi)$, and the
  mid-point of the leading edge gap inferring from several
different forms of modelling.

For doping $p$ greater than 0.2, well-defined peaks emerge below
T$_{c}$ in the $B_{1g}$ channel at a frequency roughly consistent
 with the tunneling gap.
Both Raman and tunneling data fall below rather scattered ARPES data
The agreement between Raman and tunneling results indicates that the
 $B_{1g}$ vertex renormalization is small in the overdoped regime. This
 may be the consequence of just small enough spin-fermion coupling, or the
 partial cancellation between the renormalizations due to spin fluctuations and due to phonons. The distinction with ARPES is likely due to ARPES resolution
and also, possibly, to the effects from bilayer splitting.

 For $p$ less than or equal to the optimal value,
the Raman gap values in Fig \ref{comparison}
are mostly from Ref \cite{Hewitt}, Ref \cite{Kendziora}, and Ref
\cite{Blumberg}. They fall consistently below those from
tunneling and from ARPES although there is some degree of scatter.
In regard to the findings of this paper, this implies that the
pairing more likely arises from spin fluctuations in this range
of doping. Yet the growth of intensity predicted in Figure
(\ref{full}) is not seen. This may be a result of inelastic
scattering or may be due to the presence of the pseudogap. In
addition for reasons that are not well understood, other
groups~\cite{Hackl,Sugai} find very weak $B_{1g}$ Raman
signals from underdoped samples of Bi2212.
They thus find it difficult to impossible to extract values of
$2\Delta$ in this symmetry and doping range.

Yet this is not the case for $B_{2g}$ scattering (light
orientations rotated by 45 degrees with respect to $B_{1g}$
orientations). Peaks have been shown to occur only at
temperatures below T$_{c}$ at a frequency with scales with
T$_{c}$ {\em for all dopings}\cite{Opel2000}. This apparent
contradiction between the findings in $B_{1g}$ and $B_{2g}$
requires more study and must be reconciled before firm
conclusions can be drawn about the pairing mechanism. A possible
alternative scenario could be that the anti-nodal quasiparticles
become gapped via a mechanism which is not related to
superconductivity, such as precursor spin- or charge-density wave
pairing. In this case, the difference in values of $2\Delta$ from
$B_{1g}$ Raman, ARPES, and STM and other probes would not be
unexpected. This remains very much an open question meriting
further investigation.

To conclude, in this paper, we considered $B_{1g}$ Raman
intensity in a $d-$wave superconductor. For non-interacting
fermions, $B_{1g}$ intensity is peaked at $2\Delta$. We found
that for interacting fermions, there is a qualitative distinction
between spin-mediated and phonon-mediated interactions. For
spin-mediated interaction, the peak in the intensity shifts
downwards compared to $2\Delta$ due to the development of the
resonance below $2\Delta$. We verified that the resonance
survives even if the interaction is retarded. For phonon-mediated
interaction, the resonance does not develop, and the peak remains
at $2\Delta$. The effect is quite generic and is also valid if
phonons are replaced by charge-density-waves. We presented the
generic condition under which resonance occurs in a Raman vertex
of arbitrary symmetry.

We are thankful to G. Blumberg, R. Hackl, A. Sacuto, and D. Morr
 for useful discussions.
The research was supported by NSF DMR 0240238 (A.V. Ch),
Alexander von Humboldt Foundation, ONR Grant No. N00014-05-1-0127,
and NSERC (T.P.D.).

\end{document}